\documentclass [aps,twocolumn,superscriptaddress,altaffilletter,lengthcheck,tightenlines,showpacs,showkeys]{revtex4}
\usepackage[dvipdf]{epsfig}

\newcommand{\ben}{\begin{eqnarray}}
\newcommand{\een}{\end{eqnarray}}
\newcommand{\be}{\begin{equation}}
\newcommand{\ee}{\end{equation}}
\newcommand{\ba}{\begin{eqnarray}}
\newcommand{\ea}{\end{eqnarray}}

\begin{document}

\title{Cylindrical  wormholes  with  positive cosmological constant}


\author{Mart\'{\i}n G. Richarte}\email{martin@df.uba.ar}
\affiliation{ Departamento de F\'{\i}sica, Facultad de Ciencias Exactas y
Naturales,  Universidad de Buenos Aires and IFIBA, CONICET, Ciudad
Universitaria, Pabell\'on I, 1428, Buenos Aires, Argentina}

\begin{abstract}
We construct  cylindrical, traversable  wormholes with finite radii by taking into account the cut-and-paste procedure for the case of cosmic string  manifolds with a positive cosmological constant.  Under reasonable assumptions about the equation of state of the matter located at the shell, we find that the wormhole throat undergoes a monotonous evolution provided it moves  at a constant velocity.  In order to explore the dynamical nonlinear behaviour of the wormhole throat,  we  consider that the matter at the shell is supported by   anisotropic Chaplygin gas, anti-Chaplygin gas, or a mixed of Chaplygin and anti-Chaplygin gases implying that wormholes could suffer an accelerated  expansion or  contraction but the oscillatory behavior seems to be  forbidden.
\end{abstract}

\maketitle

\section{Introduction}
Cosmic strings are  primordial objects  that  may have appeared in the early Universe; these  topological defects could arise as solitons in a grand unified theory and could be produced  as a result of  symmetry-breaking  phase transition. If the symmetry  breaking occurred after inflation, the string may survive until the present Universe \cite{Vile}.
Nowadays, cosmological observations suggest  that dark energy  and dark matter are the main components of the Universe, so  the observed amount cosmic string cannot  lead to a dominant contribution to the Universe. However,  a non-negligible fraction would still be  allowed observationally \cite{PRL1}. Because cosmic strings  are very massive objects, one could expect that the gravitational effects near them could have interesting consequences regarding the bending of  light--for instance, the gravitational lensing produced by a cosmic string  would seem to produce a double image of  elliptical galaxies with an angular separation of $1.9~ \rm{arc sec}$ when the dimensionless string tension  ${\rm G}\mu$ is around $10^{-7}$ \cite{GL}. Concerning the  typical values  of the dimensionless string tension ${\rm G}\mu$,  the new stringent constraints coming from  the Planck mission based on  the cosmic microwave background power spectrum lead to ${\rm G}\mu<1.5\times 10^{-7}$ at a $95\%$ confidence level  \cite{Planck}.

As is well known, any attempt to construct thin-shell wormholes requires the use of the cut-and-paste procedure \cite{visser1}, \cite{visser2} and works with the junction conditions associated with the gravity theory under study \cite{daris}. Cylindrical thin-shell wormholes within the context of general relativity (GR) were built, and it was found that, in most of the cases, the wormholes are supported by exotic matter, violating the energy conditions \cite{CE}, \cite{CWO}, \cite{CWH}. However, alternative gravity theories such as the DGP model or  Brans-Dicke framework are a fertile arena for trying to build  cylindrical thin shells supported by nonexotic matter, where the energy conditions can be fulfilled  by  choosing  suitably the parameters of the model, \cite{MR1}, \cite{CWH2}.

A cornerstone issue of any solution of the equations of gravitation is related to its mechanical stability. The stability of cylindrical wormholes has been thoroughly studied for the case of small perturbations preserving the original symmetry of the conﬁgurations \cite{CWO}. In doing so, it was assumed that the equations of state for the dynamic case have the same form as in the static one, and  it  was found that the throat collapses to zero radius, remains static, or expands forever, depending only on the sign of its initial velocity  \cite{CWO}.  The  stability of  wormhole geometry is a physically relevant point for knowing if these gravitational configurations could  last long enough  that their
traversability makes sense. For instance,  wormholes  constructed by gluing two copies of a global cosmic string  with or without a curvature singularity seem to behave in the same manner under linear perturbation: both kind of geometries  are not stable under radial velocity perturbations and then no oscillatory behavior is possible \cite{CWH}.

In the present work, our goal is to construct a thin-shell wormholes with cylindrical symmetry, implementing the cut-and-paste procedure on cosmic strings in spacetimes with a positive cosmological constant \cite{SB}. As  was pointed out in  Ref. \cite{SB}, the reason for studying these kinds of solutions is twofold. The cosmological observations strongly suggest  that the Universe is currently speeding up due to a nonvanishing vacuum energy (dark energy) represented by a cosmological constant $\Lambda>0$, which implies the existence of a cosmic horizon of size $\ell=1/\sqrt{\Lambda}$. Since $\Lambda$ has a very tiny value, the cosmological horizon is too large, so one might be tempted to  neglect the effect of the cosmological constant on local physics. However, the solution reported in  Ref. \cite{SB} shows that a cosmic string with $\Lambda>0$ would affect not only the global topology of the spacetime but also the local physics, indicating that the wormhole geometry will be substantially different from those cases explored in  Refs. \cite{CWO} or \cite{CWH}. For instance,  the impossibility of reaching  the spatial infinity is related to the existence of a cosmological event horizon; therefore  cylindrical wormhole geometry will have a finite radius in contrast with the asymptotically flat solutions.  We are going to analyze  several definitions for the flare-out condition, taking into account the  noncompact character of the geometry \cite{Broni}.  When considering the analysis based on  velocity linear perturbations, we obtained that wormhole's throat moves at a constant velocity, so there is no emission of gravitational waves. We will explore the nonlinear behavior of the wormhole throat and its  stability when the matter located at the shell is supported by anisotropic Chaplygin gas.  In the next section, without loss of generality, we will work with  units such that  $c=\hbar=1$, and the metric signature is $(-,+,+,+)$.
\section{Thin-shell wormholes  }
The solution of Einstein's equations, $G_{ab}+\Lambda g_{ab}=8\pi GT_{ab}$, associated with a cosmic string spacetime in  vacuum (outside the string core), is given by  
\begin{eqnarray}
\label{solu1}
ds^{2}&=&f(\rho)[dz^2-dt^2]+h(\rho)d\theta^2+g(\rho)d\rho^2, \\ 
\label{sa}
f(\rho)&=&\cos^{\frac{4}{3}}\frac{\rho\sqrt{3\Lambda}}{2} ,\\
\label{sb}
h(\rho)&=&\frac{4\delta^2}{3\Lambda}\sin^{2}\frac{\rho\sqrt{3\Lambda}}{2}\cos^{-\frac{2}{3}}\frac{\rho\sqrt{3\Lambda}}{2},\,\,g(\rho)=1,
\end{eqnarray}
where the aforesaid  metric admits three Killing vectors spanned by  $\left\langle \partial_{t},\partial_{\theta}, \partial_{z} \right\rangle$, and the parameter $\delta$ is related to the deficit angle  \cite{SB}. The solution inside the core surface ($\rho=\rho_{0}$) corresponding to an Abelian Higgs cosmic string  can be obtained from Eq. (\ref{solu1}) by simply replacing  $\delta\rightarrow 1$ and $\Lambda \rightarrow \Lambda_{d}=\Lambda+2\pi {G}\lambda \eta^4$, where $\lambda$ stands for the coupling in the Mexican potential, while $\eta$ refers to the expectation value of the complex scalar field, thus $|\Phi|=\eta$. Here,  $\delta$ includes corrections dependent on $\Lambda$, which  we may consider as coming  from the finite thickness of the string but also depend on the dimensionless string tension, ${\rm G}\mu$, and  the  core size $\rho_{0}$. The size of the string is of order 
$ 1/\eta\sqrt{\lambda}$, at least when the winding number is small, because  the metric is flat on the string axis, so we can approximate $\rho_{0}$ by its value in flat space  \cite{SB}. Taking into account  the scale of symmetry breaking, $\eta$ is small compared to the Planck scale in theories of particle physics in which cosmic strings could appear; namely,  the grand unification scale is $10^{16}{\rm GeV}$, so that $ {\rm G}\eta^2 \simeq 10^{-3}$. The latter fact indicates that is reasonable to assume that the string is small compared to the cosmic horizon. In other words, the string size goes as $\rho_{0} \simeq 1/m_{\rm Higgs}$ with  $m_{\rm Higgs}\simeq 125 {\rm Gev}$, while the observed value of the vacuum dark energy is $\Lambda \simeq 10^{-82} {\rm GeV}^{2}$ implying that  the cosmic horizon occurs at $\ell =1/\sqrt{\Lambda}\simeq  10^{41} {\rm GeV}^{-1}$. Therefore, we find that the dimensionless position of the string surface $x_{\rm core}=\rho_{0}\sqrt{3\Lambda}/2 ={\cal O}(10^{-44})$ is much smaller than the dimensionless position of the cosmic horizon $x_{\rm horizon}=\ell \sqrt{3\Lambda}/2\simeq 0.86={\cal O}(1)$. Using that $\rho_{0}\sqrt{\Lambda}<<1$ along with $\rho_{0}\sqrt{\Lambda_{d}}<<1$,  it can be shown that the string mass per unit length is approximately $\mu \simeq \pi \lambda\eta^4 \rho^2_{0}$, whereas $\delta \simeq 1-4{\rm G}\mu[1+{\rm G}\mu + 3\rho^2_{0}\Lambda/4]$. Notice that the deficit in the azimuthal angle corresponds to  $\Delta\theta=2\pi(1-\delta)\simeq 2\pi[ 4{\rm G}\mu+ 3{\rm G}\mu\rho^2_{0}\Lambda+  4{\rm G}^2\mu^2]$. 
 
The wormhole construction follows  the usual steps of the cut-and-paste procedure \cite{visser1}. Starting from a manifold $M$
described by the metric in Eq. (\ref{solu1}) with local coordinates $X^\alpha=(t,\rho,\theta,z)$, we remove the region  defined by $\rho<a$,  and we  take two copies $M^+$ and $M^-$ of the resulting manifold.Then we join them at the surface $\Sigma$  defined by $\rho=a$, so that a new geodesically complete manifold ${\cal M}=M^+\cup M^-$ is obtained. The surface $\Sigma$ is a  minimal-area hypersurface satisfying the flare-out
condition: in both sides of the new manifold, surfaces of constant $\rho$  increase their areas as one moves away from $\Sigma$;
thus one says that ${\cal M}$ presents a throat at $\rho=a$. On the  surface $\Sigma$, we define coordinates $\xi^{a}=(\tau,\theta,z$), where
$\tau$ is the proper time. Then, to allow for a dynamical analysis, we  let the radius depend on $\tau$, so that  the
surface $\Sigma$ is given by the function ${\cal H}$ which fulfills the  condition ${\cal H}(\rho,\tau)=\rho-a(\tau)=0$.
As a result of pasting the two copies of the original manifold, we  have a matter shell placed at  $\rho=a$. Its
dynamical evolution is determined by the Einstein equations projected  on $\Sigma$, that is, by the Lanczos equations \cite{daris}
\be
-[K_a^{~b}]+[K]\delta_a^{~b}=
8\pi {\rm G} S_a^{~b},
\label{e10}
\ee
where $K_a^b$ is the extrinsic curvature tensor defined by
\be
  K_{ab}^{\pm} = - n_{\gamma}^{\pm} \left. \left( \frac{\partial^2
  X^{\gamma}}{\partial \xi^a \partial \xi^b} + \Gamma_{\alpha \beta}^{\gamma}
  \frac{\partial X^{\alpha}}{\partial \xi^a} \frac{\partial
  X^{\beta}}{\partial \xi^b} \right) \right|_{\Sigma}, \label{e6}
\ee
with $n_{\gamma}^{\pm}$  the unit normals ($n^{\gamma} n_{\gamma} = 1$) to
$\Sigma$ in $\mathcal{M}$:
\be
  n_{\gamma}^{\pm} = \pm \left| g^{\alpha \beta} \frac{\partial
  \mathcal{H}}{\partial X^{\alpha}} \frac{\partial \mathcal{H}}{\partial
  X^{\beta}} \right|^{- 1 / 2} \frac{\partial \mathcal{H}}{\partial
  X^{\gamma}} . \label{e7}
\ee
The bracket  $[K_a^{~b}]$ denotes the jump ${K_a^{~b}}^+ - {K_a^{~b}}^-$ across the surface $\Sigma$, 
$[K]=g^{ab}[K_{ab}]$ is the 
trace of $[K_{ab}]$, and
$S_a^{~b} = {\rm diag} ( -\sigma, 
~p_{\theta}, ~p_{z} )$ is the surface stress-energy tensor, 
with $\sigma$ the surface energy density and $p_\theta$, $p_z$  the 
surface pressures.

In terms of these functions the components of the extrinsic curvature read 
\be
{K^\tau_{~~\tau}}^{\pm} = \pm \frac{1}{2 \sqrt{1+\dot{a}^2}}\left[ 2\ddot{a}+(\dot{a}^2+1)\frac{f'(a)}{f(a)}\right]
\label{e8a}
\ee
and
\be
{K^\theta_{~~\theta}}^{\pm} = \pm\sqrt{1+
\dot{a}^2} \frac{h'(a)}{2h(a)},\,\,\,\,\ {K^z_{~~z}}^{\pm} = \pm \sqrt{1+\dot{a}^2}\frac{f'(a)}
{2f(a)}, 
\label{e8b}
\ee
where the dot means $d/d\tau$, the prime indicates a derivative with respect to $a$, and we have used the metric on $\Sigma$, $h_{ab}={\rm diag}(-1, h(a), f(a))$ for lowering the indices. Replacing these expressions in the Lanczos equations, we obtain the surface energy density $\sigma=-S_\tau^\tau$ and the pressures $p_\theta=S_\theta^\theta$ and $p_z=S_z^z$:  
\be
  \sigma = - \frac{\sqrt{1 +\dot{a}^2}}{8 \pi {\rm G}} \left( 
  \frac{f' ( a )}{f ( a )} + \frac{h' ( a )}{h ( a )} \right),
\label{e11}
\ee
\be
p_{\theta} =  \frac{1}{8 \pi{\rm G}\sqrt{1+\dot{a}^2}} 
\left[2 \ddot{a} + 2(1+\dot{a}^2)\frac{f'(a)}{f(a)}\right] ,
\label{e12}
\ee
\be
p_{z} = \frac{1}{8 \pi{\rm G}\sqrt{1+\dot{a}^2}} 
\left[2 \ddot{a} + (1+\dot{a}^2)\left(\frac{f'(a)}{f(a)}+\frac{h'(a)}{h(a)}\right)\right]  .
\label{e13}
\end{equation}
\subsection{Linear stability analysis: Energy conditions}
From the above equations, we find that for the static situation $\dot a=\ddot a=0$, the pressures and the surface energy density satisfy the equations of state
\be
\label{eos1}
p_z=-\sigma,~~~~~ p_\theta=-2\sigma\frac{f'(a)h(a)}{[f(a)h(a)]'}.
\ee
If we are interested in small velocity perturbations,  it is licit to assume that the evolution of the matter on the shell can be described as a succession of static states. Thus we shall accept, as done before  \cite{CWO, CWH}, that the form of the equations of state corresponding to  the static case is kept valid in the dynamical evolution \cite{asum}. With this assumption, the equations above lead to the equation of motion $\ddot{a}=0$, and  then $a(\tau)=a_{0}+\dot {a}_0 (\tau-\tau_{0})$, where $a_0$ and $\dot {a}_0$ are, respectively, the initial wormhole radius and its initial velocity. This shows that the sign of the velocity is determined by its initial sign, that is, after a small velocity perturbation, the throat undergoes a monotonous evolution; no oscillatory behaviour exists, at least under the approximations assumed.

Let us begin by examining the energy conditions \cite{EC} in the nonstatic case. In the case of thin-shell wormholes the radial pressure $p_r$ is zero,  and within Einstein gravity, the surface energy density must fulfill $\sigma < 0$, so the weak energy condition will be violated, but the sign of $\sigma+p_{l}$, where $p_l$ is the transverse pressure is not fixed and  depends on the values of the parameters of the system. Then, the null energy condition reads $\sigma+p_{z}>0$ along with $\sigma+ p_{\theta}>0$. Using Eqs. (\ref{e11}-\ref{e13}), we find that $\sigma+p_{z}=2\ddot{a}[8 \pi{\rm G}\sqrt{1+\dot{a}^2}]^{-1}$ and $\sigma+ p_{\theta}=[2\ddot{a}-(1+\dot{a}^2)(h'/h)][8 \pi{\rm G}\sqrt{1+\dot{a}^2}]^{-1}$; then,  in order to fulfill the null energy condition we must admit only wormholes with  $\ddot{a}>0$ along with  $\ddot{a}\geq (1+\dot{a}^2)(h'/2h)>0$. For static wormholes, we obtain that $\sigma+p_{z}=0$ and $\sigma+ p_{\theta}=-(h'/h)/(8 \pi{\rm G})<0$, so it is clear that the  null energy condition is not fulfilled. 

\subsection{Nonlinear dynamic for a Chaplygin gas}
Now, we are going to drop the assumption of linear perturbations, throwing away the use of the equations of state given by  Eq. (\ref{eos1}) for studying the dynamic of the wormhole's throat. Instead of that,  we shall assume that  matter located at the shell  has  an equation of state  corresponding to a Chaplygin gas \cite{kami1}, \cite{kami2}- namely,  the pressures and the surface energy density satisfy  the relations 
\be
\label{eos2}
p_z=-\frac{{\cal A}_{z}}{|\sigma|},~~~~~ p_\theta=-\frac{{\cal A}_{\theta}}{|\sigma|},
\ee
where ${\cal A}_{i}$ is a positive constant for the case of Chaplygin gas and a negative  one for anti-Chaplygin gas \cite{kami2}.  

Combining Eqs. (\ref{e11}-\ref{e13}) along with Eq. (\ref{eos2}),    we obtain the analogous problem of  a particle in a one-dimensional potential, $\dot{x}^2(t)+{\cal V}(x)=0$, where we  have introduced some dimensionaless variables,  $x=\alpha a$ and  $t=\alpha \tau$, with $\alpha=\sqrt{3\Lambda}/2$. As a result of the previous procedure,  the potential can be rewritten as follows:
\be
\label{VP}
{\cal V}(x)=1+ \frac{\Omega~\sin^2 x}{ [3-\tan^2x]},
\ee
where $\Omega=(8\pi{\rm G})^2({\cal A}_{z}-{\cal A}_{\theta})/3\Lambda^2$ is a dimensionless parameter and its sign is controlled by the difference $\Delta{\cal A}={\cal A}_{z}-{\cal A}_{\theta}$. It turns out that Eq. (\ref{VP}) is negative definite for $x\in (x_{\rm core}, x_{\rm horizon})$ if $\Omega\leq\Omega_{c}\simeq -10^{10}$ . In order to continue the analysis of the dynamic of the wormhole, we need to solve  the energy conservation equation,  which we will tackle next. Concerning this aim, it is convenient to use  $dx=\dot{x}dt$:
\be
\label{In}
\pm(t-t_{0})=\int^{x(t)}_{x(t_{0})}{\frac{dx}{\sqrt{-{\cal V}(x)}}}.
\ee
We numerically integrate Eq. (\ref{In}) for $\Omega=-10^{10}$; in doing so we take $x \in [0.1, 0.8]$ to be far away from the string core and the cosmic horizon as well. For the case of Eq. (\ref{In})  with $(+)$, we find that the wormhole throat  exhibits a monotonous evolution with $\dot{x}>0$ and $\ddot{x}=-{\cal V}'/2>0$: the wormhole undergoes an accelerated expansion.  For the sake of completeness, we mention the reversal  situation  with $(-)$ in Eq. (\ref{In}): the wormhole undergoes an accelerated contraction, provided $\dot{x}$ grows in absolute value while $\ddot{x}$ takes  negative values. The aforesaid facts indicate that wormhole throats are unstable, expanding or skrinking in an accelerated manner.

Let us make a comment  concerning the matter located at the wormhole throat. In order for the potential to be well definite, it  is needed that  $\Omega<0$. This condition can be fulfilled in three different ways: (i) the wormhole is supported by Chaplygin gases so that  ${\cal A}_{z}>0$ and  ${\cal A}_{\theta}>0$ along with ${\cal A}_{\theta}>{\cal A}_{z}$; (ii) the wormhole is supported by anti-Chaplygin gases  so that  ${\cal A}_{z}<0$ and  ${\cal A}_{\theta}<0$ along with $|{\cal A}_{z}|>|{\cal A}_{\theta}|$;
(iii) the wormhole admits  the Chaplygin equations of state for the angular pressure [${\cal A}_{\theta}>0$] and admits for the axis of symmetry   ${\cal A}_{z}<0$, provided ${\rm sign}[\Omega]={\rm sign}[-|{\cal A}_{z}|-|{\cal A}_{\theta}|]<0$, thus $|{\cal A}_{z}|>-|{\cal A}_{\theta}|$.
\section{Summary and discussion}

In this paper, we have built  cylindrical thin-shell wormholes by gluing two copies of a cosmic string's manifold endowed with a positive cosmological constant \cite{SB}. We have shown that wormhole geometries do not exhibit an asymptotic flatness behavior at large radii  because the cosmological constant  implies the existence of a cosmic horizon of size $\ell=1/\sqrt{\Lambda}$; thus the global topology of  wormhole-like geometries with $\Lambda>0$ enforces  that such gravitational configurations must have finite radii always.

Taking into account the standard junction conditions within the context of general relativity  along with reasonable assumptions about the equation of state of the matter located at the shell,  we have shown that the wormhole throat undergoes a monotonous evolution provided it moves  at a constant velocity for the case of linear perturbations preserving the cylindrical symmetry. Leaving aside the linear radial perturbations, we have examined the full nonlinear behavior of the wormhole throat, assuming that the matter supporting these configurations is a  Chaplygin or anti-Chaplygin gas for the pressure along the string axis or in the azimuthal direction; thus, $p_{i}={\cal A}_{i}/\sigma$ with $ {\rm i}=\{ z,\theta\}$. In doing so, we have found that the wormhole geometries could exhibit an accelerated  expansion or contraction depending on the relation  between the cosmological constant and the Chaplygin gas coupling ${\cal A}_{i}$. But most importantly,  it turned out  to be that the sign of  difference $ {\cal A}_{z}-{\cal A}_{\theta}$ played  a central role in analyzing the dynamic of the  wormhole throat, and thus for integrating numerically the projected field equation as well, and determinig what kind of equations of state could admit the tangential pressures, namely, $p_z$ and $p_{\theta}$ respectively. Indeed, we have found that wormholes could be supported by pressures $p_z(\sigma)$ and $p_{\theta}(\sigma)$ with (i)  Chaplygin-Chaplygin equations of state along with  the condition   ${\cal A}_{z}<{\cal A}_{\theta}$; (ii)  anti-Chaplygin- Chaplygin equations of state along with  the condition   $|{\cal A}_{\theta}|<|{\cal A}_{z}|$; and finally (iii) a mix of anti-Chaplygin and Chaplygin gases with $|{\cal A}_{z}|>-|{\cal A}_{\theta}|$. 

Finally, we discuss  the different notions of the flare-out condition examined in the literature \cite{Broni} for the wormhole geometry with a positive cosmological constant [Eq. \ref{solu1}].  For instance,   the area per unit length is given by  ${\cal A}(a)/\ell_{z}=2\pi\delta \sqrt{f(a)h(a)}$, so it is an increasing function on both side of the throat because ${\cal A}'(a) /\ell_{z}=(8\pi^2\delta^4/\sqrt{9\Lambda})  \sin(a\alpha)\cos^{5/3}(a\alpha)[3-\tan^{2}(a\alpha)]/ {\cal A}>0$ for $\alpha a \in  (0.1; 0.8)\subset (x_{\rm core}; x_{\rm horizon} )$. Besides, the topology of the  throat with $z$ fixed corresponds to  a circle with a radius function  ${\cal R}(a)=2\pi\delta \sqrt{h(a)}$ that defines its perimeter; this is indeed an increasing function provided ${\cal R}'(a)>0$ for $a \in (a_{\rm core}; a_{\rm horizon} )$. We have shown that both the standard areal surface and  the circular radius function  reach a minimum at the position of the throat, but the latter one is less restrictive than the former one.

\acknowledgments
We are grateful to the referee for his valuable comments that helped improve the article.
The author is partially supported by the Postdoctoral
Fellowship Programme of  Consejo de Investigaciones Cient\'{\i}ficas y T\'ecnicas (CONICET).


\begin{thebibliography}{99}
\bibitem{Vile}
A. Vilenkin and E. P. S. Shellard, \emph{Cosmic Strings and Other Topological Defects} (Cambridge
University Press, Cambridge, 1994).
\bibitem{PRL1}
J.Rocher and M. Sakellariadou, Phys. Rev. Lett. {\bf 94}, 011303 (2005).
\bibitem{GL}
M.Sazhin \emph{et al.}, Mon. Not. R. Astron. Soc. {\bf 343}, 353 (2003).
\bibitem{Planck}
P.A.R. Ade \emph{et al.}, [arxiv:1303.5085].
\bibitem{visser1} M. Visser, \textit{Lorentzian Wormholes} (AIP Press, New
York, 1996). 
\bibitem{visser2}	
J. P.S. Lemos and  F. S.N. Lobo,  Phys.Rev. D {\bf 69} (2004) 104007;
 M. G. Richarte and C. Simeone , Int.J.Mod.Phys.D {\bf 17}, 1179-1196,2008
M. G. Richarte and C. Simeone, Phys. Rev. D {\bf 76}, 087502 (2007); Erratum-ibid {\bf 77}, 089903 (2008);
M. G. Richarte and C. Simeone, Phys.Rev. D {\bf 80},  104033 (2009), Erratum-ibid. D {\bf 81},  109903 (2010); 
E. F. Eiroa, M. G. Richarte, and C. Simeone,  Phys.Lett.A {\bf 373}, 1-4 (2008), Erratum-ibid.373:2399-2400, 2009; 
Goncalo A.S. Dias and Jose P.S. Lemos, Phys.Rev. D {\bf 82} (2010) 084023.
M. G. Richarte, Phys.Rev.D {\bf 82}, 044021 (2010).
M. G. Richarte, C. Simeone, Int.J.Mod.Phys.D {\bf 17}, 1179-1196 (2008);
Jose P.S. Lemos , Francisco S.N. Lobo,  Phys.Rev. D {\bf 78} (2008) 044030;  
K. A. Bronnikov, A. A. Starobinsky, Mod.Phys.Lett. A{\bf 24} (2009) 1559-1564. 
\bibitem{daris}  G. Darmois, M\'{e}morial des
Sciences Math\'{e}matiques, Fascicule XXV ch V (Gauthier-Villars, Paris, 1927)
; W. Israel, Nuovo Cimento \textbf{44B}, 1 (1966); \textbf{48B}, 463(E) (1967).
\bibitem{CWO}
E. F. Eiroa, C. Simeone,  Phys.Rev. D {\bf 70} (2004) 044008. 
\bibitem{CE}
C.Simeone,  Int.J.Mod.Phys. D {\bf 21} (2012) 1250015. 
\bibitem{CWH}
C. Bejarano, E. F. Eiroa, C. Simeone, Phys.Rev. D
{\bf 75} (2007) 027501; 	
E. F. Eiroa , C. Simeone,  Phys.Rev. D {\bf 81} (2010) 084022; 
M. G. Richarte, C.Simeone,  Phys.Rev. D {\bf 79}, 127502  (2009)  .  
Rub\'in de Celis, E.; Santill\'an, O. P.; Simeone, C.  Phys.Rev. D {\bf 86} (2012)  124009.
Sharif, M.; Azam, M., JCAP {\bf 04},  023 (2013).
\bibitem{MR1}
M. G. Richarte, Phys.Rev.D {\bf 87}, 067503 (2013).
\bibitem{CWH2}
E. F. Eiroa,  C. Simeone,  Phys.Rev. D {\bf 82} (2010) 084039. 
\bibitem{SB}
S. Bhattacharya and A. Lahiri,  Phys.Rev. D {\bf 78}, 065028  (2008).
\bibitem{Broni}
K.A. Bronnikov and J.P.S. Lemos, Phys. Rev. D {\bf 79}, 104019 (2009).
\bibitem{asum}
The perturbative treatment avoids possible subtleties regarding the validity of the static geometry for $\rho>a$ in the presence of a cylindrical moving shell.
\bibitem{kami1}
V. Gorini; A. Kamenshchik; U. Moschella,
 Phys.Rev. D {\bf 67},  063509 (2003).
\bibitem{kami2}
L.P. Chimento, Phys.Rev. D {\bf 69}   123517 (2004)
\bibitem{EC}
The \emph{weak energy condition} (WEC) states that for any timelike vector $u^{\alpha}$, it must be $T_{\alpha\,\beta}u^{\alpha}u^{\beta}\geq 0$. The WEC also implies, by continuity, the \emph{null energy condition} (NEC), which means that for any null
vector $k^{\alpha}$, it must be $T_{\alpha\,\beta}k^{\alpha}k^{\beta}\geq 0$ \cite{visser1}. In an orthonormal basis, the WEC reads $\rho\geq 0$, $\rho + p_{l}\geq 0$ $\forall ~ l$,  while the NEC takes the form $\rho + p_{l}\geq 0$ $\forall ~ l$. 
\end{thebibliography}
\end{document}